\documentclass[5p]{elsarticle}
\usepackage{graphicx}
\usepackage{amsmath,amssymb}
\usepackage[dvipdfm, colorlinks=true, pdfstartview=FitV, linkcolor=red, citecolor=blue, urlcolor=blue]{hyperref}
\usepackage{color}
\newcommand{\vect}[1]{\mathbf{#1}}

\newcommand{\im}{\mathrm{Im}\,}
\newcommand{\re}{\mathrm{Re}\,}
\newcommand{\sgn}{\mathrm{sgn}}
\newcommand{\Slash}[1]{\ooalign{\hfil/\hfil\crcr$#1$}}
\newcommand{\vzero}{\vect{0}}
\newcommand{\vp}{\vect{p}}

\newcommand{\vgamma}{{\boldsymbol \gamma}}

\newcommand{\pw}{(\omega,\vp)}
\newcommand{\GR}{G^R}
\newcommand{\DR}{D^{R}}

\newcommand{\GF}{G^{S}}
\newcommand{\DF}{D^{S}}

\newcommand{\mb}{m_b}
\newcommand{\mf}{m_f}

\newcommand{\cp}{g}
\newcommand{\kernel}{K}
\newcommand{\selfEnergyR}{\Sigma^R}

\begin{document}
\title{Ultrasoft Fermionic Mode in Yukawa Theory at High Temperature}

\author[riken,kyoto]{Yoshimasa Hidaka}
\ead{hidaka@riken.jp}

\author[kyoto]{Daisuke Satow}
\ead{d-sato@ruby.scphys.kyoto-u.ac.jp}

\author[kyoto]{Teiji Kunihiro}
\ead{kunihiro@ruby.scphys.kyoto-u.ac.jp} 

\address[riken]{Mathematical Physics Laboratory, RIKEN Nishina Center, Saitama 351-0198, Japan}
\address[kyoto]{Department of Physics, Faculty of Science, Kyoto University, Kitashirakawa Oiwakecho, Sakyo-ku, Kyoto 606-8502, Japan}

\begin{abstract}
We explore whether an ultrasoft fermionic mode exists at extremely high temperature in Yukawa theory 
with massless fermion (coupling constant is $\cp$).
We find that the fermion propagator has a pole at $\omega = \pm |\vp|/3-i\zeta$ 
for ultrasoft momentum $\vp$, where $\zeta \sim \cp^4T \ln \cp^{-1}$,
and the residue is $Z\sim \cp^2$.  It is shown that one  needs to take into account 
the  asymptotic masses and the damping rate of hard particles to get a
sensible result for such an ultrasoft fermionic mode; 
possible  vertex correction turns out unnecessary 
for the scalar coupling in contrast to the gauge coupling. 
\end{abstract}

\date{\today}

\maketitle

\section{Introduction}
\label{sec:introduction}
\begin{figure}
\begin{center}
\includegraphics[width=0.4\textwidth]{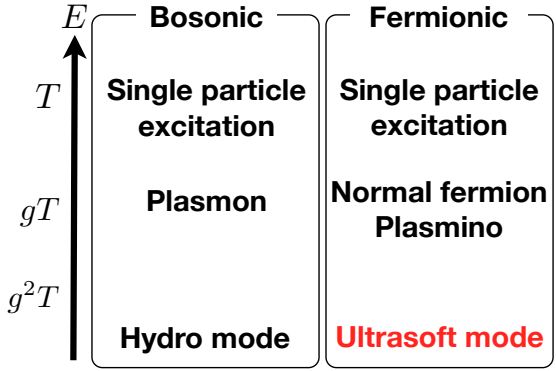}
\caption{Schematic illustration of the energy hierarchy of
bosonic and fermionic modes in a massless fermion-boson system at finite temperature:
The vertical axis is the energy scale.
The left  and right part display the cases of  bosonic and  fermionic mode, respectively.
The bosonic modes may be of scalar-, gauge-bosonic and hydrodynamic ones.
No established fermionic mode had been known in the ultrasoft region.  
}
\label{fig:collective}
\end{center}
\end{figure}

It is of basic importance to clarify the nature of  quasiparticle or
collective excitations for understanding a many-body system,
in particular, in the low-energy regime.
In a fermion-boson system such as  quantum electrodynamics (QED), quantum chromodynamics (QCD), and Yukawa theory
at so high temperature $T$ that the masses of the particles are negligible, 
the average inter-particle distance is proportional to $1/T$ and some collective effects can be expected 
in the soft momentum scale $1/\cp T$ with $\cp$ being the coupling constant.
Indeed, soft bosonic modes in the longitudinal as well as the transverse channels
exist and are known as plasmon~\cite{plasmon}, 
while the fermionic counter part is known as plasmino~\cite{HTL,plasmino}, 
both of which have masses of order $\cp T$.
In the lower energy region, there exist 
hydrodynamic modes of bosonic nature, which are actually the  zero modes 
associated with the conservation of energy-momentum and charges.
The energy hierarchy of bosonic and fermionic modes at high temperature 
may be summarized schematically as shown in Fig.~\ref{fig:collective}.
From the figure, one may have an intriguing but natural question whether such ultrasoft 
($\lesssim \cp^2T$) or zero modes can exist also in the fermion sector, 
possibly when the fermion system has a peculiar symmetry
 such as a chiral symmetry. 

In this article, we argue and demonstrate that such a fermion mode can exist in this infrared energy region.
Here we should mention that there have been some suggestive works for supporting the existence of 
such an ultrasoft fermionic mode at finite temperature.
It was shown in a one-loop calculation \cite{3peak,mitsutani} 
that when a fermion is coupled with a massive boson with  mass $m$, 
the spectral function of the fermion gets to have  a novel peak in the far-low-energy region  in addition 
to the normal fermion and the plasmino, when $T\sim m$,
 irrespective of the type of boson; it means that 
 the spectral function of the fermion has a three-peak structure in this temperature region. 
Very recently, the present authors \cite{shk} have suggested that such a three-peak
structure may persist even at the high temperature limit in the sense $m/T\rightarrow 0$, for the massive vector boson on the basis of a gauge-invariant formalism, again, at the one-loop order. 
Thus, one may expect that the novel excitation may exist in the far-infrared region 
also for a fermion coupled with a massless boson,
although the one-loop analysis admittedly may not be applicable at the ultrasoft momentum region.

In supersymmetric models, the ultrasoft modes are discussed as Nambu-Goldstone fermions called phonino associated with
spontaneously breaking of supersymmetry at finite temperature, which is shown by using Ward-Takahashi identity and a diagrammatic technique~\cite{phonino,lebedev1}. The discussion is extended to QCD at weak coupling~\cite{lebedev2}, in which a supersymmetry is assigned at  the vanishing coupling. The supersymmetry is explicitly broken by interactions; thus the phonino is not the exact zero mode but a pseudo-phonino.

In this article, we analyze the fermion propagator in the ultrasoft energy region in Yukawa theory, which is a nonsupersymmetric model,
because it is the simplest fermion-boson system and hence a good theoretical laboratory.
We discuss the existence and the properties of the fermionic 
mode in the ultrasoft region using a similar diagrammatic technique in Refs.~\cite{lebedev1,lebedev2, hidaka}.
We deal with a massless fermion. In this case, the important point is that the fermion is chiral;
the real part of the retarded fermion self-energy, $\Sigma^R(\omega,\vzero)$, 
at zero spatial momentum and the vanishing chemical potential,
is an odd function of $\omega$:
\begin{align}
\re\Sigma^R(-\omega,\vzero)=-\re\Sigma^R(\omega,\vzero).
\end{align} 
If the $\re\Sigma^R(\omega,\vzero)$ is  a smooth function at $\omega=0$, then,
 $\re\Sigma^R(-\omega,\vzero)=0$ at $\omega=0$, 
which implies $\re G^{-1}(\omega,\vzero)=0$ at $\omega=0$;
the spectrum has a peak at the origin provided that the imaginary part of the fermion self-energy 
is not too large.
This argument suggests that the existence of the ultrasoft pole may be a universal phenomenon
 at high temperature in the theory composed of  massless fermions coupled with a boson, 
which includes not only Yukawa theory but also QED and QCD.

It is, however,  not a simple task to establish that fermionic modes exist in the ultrasoft region on a general ground beyond the one-loop order 
because of the infrared divergence called pinch singularity 
\cite{lebedev2,hidaka, transport,ultrasoft-am} 
that breaks a naive perturbation theory, as will be briefly reviewed 
in the next section.
We remark that the same difficulty arises in the calculation of transport coefficients 
\cite{hidaka,transport} and the gluon self-energy \cite{ultrasoft-am}.

We shall show that 
the retarded fermion propagator has a pole 
at $\omega = \pm |\vp|/3-i\zeta$  ($\zeta\sim \cp^4T\ln \cp^{-1}$ is introduced later)
with the residue $Z\sim \cp^2$ for ultrasoft momentum $\vp$. 

\begin{figure}
\begin{center}
\includegraphics[width=0.26\textwidth]{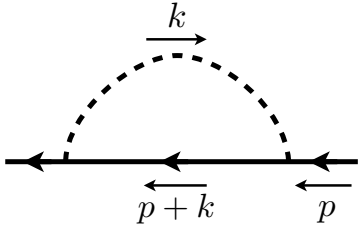}
\caption{Diagrammatic representation of the fermion self-energy in Eq.~(\ref{eq:one-loop-bare}) 
at one-loop order.
The solid and dashed line are the propagator of the fermion and the scalar boson, respectively.
In Eq.~(\ref{eq:one-loop-pc}), the fermion and the boson propagator in the internal lines 
are replaced by the dressed propagators given in Eq.~(\ref{eq:resum-propagator}).}
\label{fig:oneloop}
\end{center}
\end{figure}

\section{Ultrasoft Fermionic Mode}
\label{sec:resum}

As stated in the Introduction,
we deal with a massless fermion (denoted by $\psi$) coupled with a
scalar field $\phi$ through the interaction Lagrangian 
${\cal L}_I=g\bar{\psi}\psi\phi$ at high temperature $T$.
We do not take into account  the self-coupling of the scalar fields for simplicity.
In this section, we calculate the fermion self-energy to
obtain the fermion retarded Green function 
with an ultrasoft momentum $p\lesssim \cp^2T$
at high temperature.
We first see that the naive perturbation theory breaks down
in this case. 
Then, we shall show that a use of a dressed propagator gives
a sensible result in the perturbation theory 
and that  the resulting fermion propagator has a new pole in the ultrasoft region.

The retarded self-energy in the one-loop level is given 
by
\begin{equation}
\label{eq:one-loop-bare}
\begin{split}
\selfEnergyR_{\text{bare}}(p)&= i\cp^2\int\frac{d^4k}{(2\pi)^4}\Bigl[\DF_0(-k)\GR_0(p+k)\\
&\qquad\qquad\qquad\qquad+\DR_0(-k)\GF_0(p+k)\Bigr],
\end{split}
\end{equation}
where $D_0^{R,S}(-k)$ and $G_0^{R,S}(p+k)$ are the {\it bare} propagators of the fermion and the scalar boson defined as
\begin{align}
\label{eq:bare-propagatorS}
\GR_0(k)&=\frac{-\Slash{k}}{k^2+ik^0\epsilon},\\
\GF_0(k)&=\left(\frac{1}{2}-n_F(k^0)\right)i\Slash{k}(2\pi)\sgn(k^0)\delta(k^2),
\\
\DR_0(k)&=\frac{-1}{k^2+ik^0\epsilon},\\
\DF_0(k)&=\left(\frac{1}{2}+n_B(k^0)\right)i(2\pi)\sgn(k^0)\delta(k^2).
\label{eq:bare-propagatorE}
\end{align}
Here, $n_F(k^0)\equiv1/(\exp(k^0/T)+1)$ and $n_B(k^0)\equiv {1/(\exp(k^0/T)-1)}$ are 
the Fermi-Dirac and Bose-Einstein  distribution functions, respectively.
In the present analysis, we have preferentially employed
 the real-time formalism in Keldysh basis~\cite{Keldysh}. 
The diagrammatic representation of Eq.~(\ref{eq:one-loop-bare}) is shown in Fig.~\ref{fig:oneloop}. 
Inserting Eqs.~(\ref{eq:bare-propagatorS}) through (\ref{eq:bare-propagatorE})
into (\ref{eq:one-loop-bare}), we obtain
\begin{equation}
\begin{split}
\selfEnergyR_{\text{bare}}(p)&= \cp^2\int\frac{d^4k}{(2\pi)^4}\Bigl[\left(\frac{1}{2}+n_B(k^0)\right)\\
&\quad\times\frac{\Slash{k}+\Slash{p}}{p^2+2p\cdot k+i(k^0+p^0)\epsilon}(2\pi)\sgn(k^0)\delta(k^2)\\
&\qquad-\left(\frac{1}{2}-n_F(k^0+p^0)\right)
\frac{\Slash{k}+\Slash{p}}{p^2+2p\cdot k+ik^0\epsilon}\\
&\qquad\qquad\times(2\pi)\sgn(k^0+p^0)\delta((k+p)^2)\Bigr],
\end{split}
\end{equation}
where we have used the on-shell conditions for the bare particles, 
$k^2=0$, and $(k+p)^2=0$ in $\GF_0(p+k)$ and $\DF_0(-k)$.
Then, for small $p$, the self-energy is reduced to
\begin{equation}
\begin{split}
\selfEnergyR_{\text{bare}}(p)&= \cp^2\int\frac{d^4k}{(2\pi)^4}\kernel(k)\frac{\Slash{k}}{2p\cdot k+ik^0\epsilon},
\end{split}
\end{equation}
where $\kernel(k)=(2\pi)\sgn(k^0)\delta(k^2)(n_F(k^0)+n_B(k^0))$, which is independent of $p$.
This approximation is equivalent to the HTL approximation~\cite{HTL}.
The HTL approximation is, however, only valid for
$p\sim \cp T$, and not applicable in the ultrasoft momentum region.
In fact, the retarded self-energy in the one-loop level obtained with use of the bare
propagators is found to diverge when $p\rightarrow 0$,
since the integrand contains $1/ p\cdot k$,
which singularity is the so called ``pinch singularity'' \cite{lebedev2,hidaka,transport,ultrasoft-am}.

The origin of this singularity is traced back to the use 
of the bare propagators because the singularity is caused by the fact that the dispersion 
relations of the fermion and the boson are the same and the damping rates are zero in that propagators.
For this reason, one may suspect that this singularity can be removed
by adopting the dressed propagators taking into account the asymptotic masses and decay width 
of the quasiparticles,
as will be shown shortly.

Noting that the leading contribution comes from the hard ($k \sim T$) internal and almost on-shell ($k^2\approx 0$)
momentum\footnote{Here we justify neglecting the case that the internal momenta are soft ($k\sim gT$) or smaller;
for the soft momentum, the HTL resummed propagators \cite{HTL-resum} should be used.
In these propagators, the dispersion relations of the fermion and the boson are different, so the pinch singularity does not appear.},
we are lead to employ the following dressed propagators for the fermion and boson:
\begin{align}
\label{eq:resum-propagator}
\GR(k)\simeq&-\frac{\Slash{k}}{k^2-m^2_f+2i\zeta_f k^0},\\
\GF(k)\simeq&\left(\frac{1}{2}-n_F(k^0)\right)\Slash{k}\frac{4i\zeta_fk^0}{(k^2-m^2_f)^2+4\zeta^2_f(k^0)^2},\\
\DR(k)\simeq&-\frac{1}{k^2-m^2_b+2i\zeta_b k^0},\\
\DF(k)\simeq&\left(\frac{1}{2}+n_B(k^0)\right)\frac{4i\zeta_bk^0}{(k^2-m^2_b)^2+4\zeta^2_b(k^0)^2} ,
\end{align}
where $\mf^2\equiv \cp^2 T^2/8$ and $\mb^2\equiv \cp^2 T^2/6$ are the so called asymptotic masses  
of the fermion and the boson at $k^2\simeq0$, respectively~\cite{scalar,dispersion-numerical,Blaizot:2001nr}.
The damping rates of the hard particles, $\zeta_f$ and $\zeta_b$, are of order $\cp^4 T\ln \cp^{-1}$. The logarithmic enhancement for the damping rate of the scalar particle is caused by the soft-fermion exchange, which is
the similar enhancement to that of the hard photon~\cite{Kapusta,Baier}.
Using these dressed propagators, we have
\begin{equation}
\label{eq:one-loop-pc}
\Sigma^R(p)\simeq \cp^2\int\frac{d^4k}{(2\pi)^4}\kernel(k)
\frac{\Slash{k}}{2p\cdot k +2i k^0\zeta+ \delta m^2 }
\end{equation}
for small $p$,
where $\delta m^2\equiv \mb^2-\mf^2=\cp^2T^2/24$ and $\zeta\equiv \zeta_f+\zeta_b$. 
Here we have used the modified on-shell condition of  the quasi-particles,
$k^2-m_f^2+2i\zeta_f k^0=0$ and $k^2-m_b^2+2i\zeta_f b^0=0$, 
to obtain the denominator of the integrand in Eq.~(\ref{eq:one-loop-pc}).
We neglected $m_b$, $m_f$, $\zeta_b $, and $\zeta_f $ in $\kernel(k)$,
since the leading contribution comes from hard momenta $k\sim T$.
It is worth emphasizing that thanks to $\delta m^2$ and $\zeta$, 
$\Sigma^R(p)$ given in Eq.~(\ref{eq:one-loop-pc}) does not diverge in the infrared limit, $p\rightarrow 0$.

\begin{figure}
\begin{center}
\includegraphics[width=0.35\textwidth]{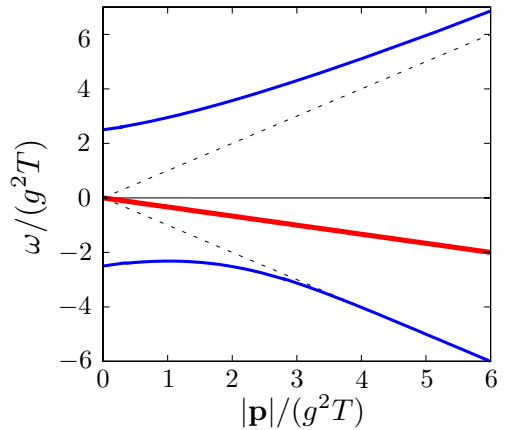}
\caption{The dispersion relation in the fermionic sector.
The coupling constant is set to $\cp=0.1$.
The vertical axis is the energy $\omega$, while the horizontal axis is the momentum $|\vp|$.
The solid (blue) lines correspond to the normal fermion and the antiplasmino, while the bold solid (red) one to the ultrasoft mode.
Notice that since we focus on the fermion sector, the antiplasmino appears instead of the plasmino.
The dotted lines denote the light cone.
Since our analysis on the ultrasoft mode is valid only in $|\vp|\ll \cp^2T$, 
the plot in $|\vp|\gtrsim \cp^2T$ may not have physical meaning.
The residue of the antiplasmino becomes exponentially small 
for $|\vp|\gg \cp T$, 
so the plot of the antiplasmino does not represent physical excitation in $|\vp|\gg \cp T$, either.
}
\label{fig:dispersion}
\end{center}
\end{figure}

Here, we expand Eq.~(\ref{eq:one-loop-pc}) in terms of $2(p\cdot k+i k^0\zeta)/\delta m^2$.
The leading contribution reads
\begin{equation}
\label{eq:sigma-p1}
\begin{split}
\Sigma^R(p)&\simeq-\cp^2\int \frac{d^4k}{(2\pi)^4}K(k)\Slash{k}\frac{2ik^0\zeta+2p\cdot k}{(\delta m^2)^2}\\
&=-\frac{1}{Z}\left((p^0+i\zeta)\gamma^0+v \vp\cdot \vgamma\right),
\end{split}
\end{equation}
with $Z\equiv 8(\delta m^2)^2/(\cp^2\pi^2T^4)$ and $v=1/3$.
Thus, we get the fermion propagator in the ultrasoft region as
\begin{equation}
\label{eq:result-propagator}
\begin{split}
\GR\pw=&-\frac{1}{\Slash{p}-\Sigma^{\text {R}}\pw}
\simeq\frac{1}{\Sigma^{\text {R}}\pw}\\
=&-\frac{Z}{2}\left(\frac{\gamma^0-\hat{\vp}\cdot\vgamma }{\omega+v|\vp|+i\zeta}+\frac{\gamma^0+\hat{\vp}\cdot\vgamma}{\omega-v|\vp|+i\zeta}\right).
\end{split}
\end{equation}
Here we have decomposed the fermion propagator into the fermion number $+1$ and $-1$ sectors
in the second line.
These two sectors are symmetric under $\vp\leftrightarrow -\vp$ and $v\leftrightarrow -v$,
so we analyze only the fermion number $+1$ sector in the following:
From Eq.~(\ref{eq:result-propagator}), we find a new pole at 
\begin{align}
\label{eq:result-dispersion}
\omega=- v|\vp|+i\zeta.
\end{align}
Notice that the real part of $\omega$ is negative for the fermion sector, which
 suggests that this peak has an antifermion-hole-like character
like the antiplasmino \cite{plasmino}.
The dispersion relation of the real part, $\re\omega=-v|\vp|$, is shown 
in Fig.~\ref{fig:dispersion} together with the HTL results \cite{HTL,plasmino} for comparison,
where the coupling constant is chosen as $\cp=0.1$.
The imaginary part of the pole reads
\begin{equation}
\label{eq:result-width}
\zeta \sim \cp^4T \ln \cp^{-1},
\end{equation}
which is much smaller than those of the normal fermion and the antiplasmino \cite{scalar}.
Since the real part and the imaginary part of the pole are finite for $|\vp|\neq 0$, 
this mode is a damped oscillation mode. 
The residue at the pole is evaluated to be
\begin{align}
\label{eq:result-residue}
Z =\frac{\cp^2}{72\pi^2}\sim \cp^2,
\end{align}
which means that the mode has only a weak strength in comparison with those of the normal fermion 
and the antiplasmino, 
whose residues are order of unity.
It is worth mentioning that 
such a smallness of the residue is actually compatible with the results in the HTL approximation:
The sum of the residues of the normal fermion and the anti-plasmino modes obtained
in the HTL approximation is unity and thus the sum rule of the spectral function of the fermion
is satisfied in the leading order.
Therefore, one could have anticipated that the residue of the ultrasoft mode 
can not be the order of unity but should be of higher order.

Equations~(\ref{eq:result-propagator}) through (\ref{eq:result-residue}) for Yukawa theory
are obtained for the first time and  constitute the main results 
of the present work. We emphasize again that the usual HTL approximation \cite{HTL,plasmino}
is inapplicable in this ultrasoft energy region, and hence such an ultrasoft 
peak in the spectral function of the fermion necessarily has never been found before. 
As a summary, the respective coupling orders of the spectral properties of the ultrasoft mode 
are tabulated in Table.~\ref{tab:result}, together with those of
the normal fermion and the antiplasmino obtained in the HTL approximation.

\begin{figure}
\begin{center}
\includegraphics[width=0.4\textwidth]{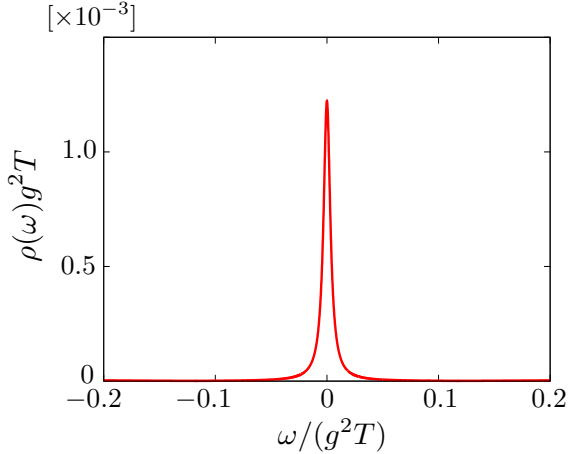}
\caption{The spectral function in the fermion sector, Eq.~(\ref{eq:spectral}), as a function of energy $\omega$ at zero momentum.
The coupling constant is set to $\cp=0.1$.}
\label{fig:spectle}
\end{center}
\end{figure}

The pole given by Eq.~(\ref{eq:result-dispersion}) gives rise to a new peak in the spectral 
function of the fermion as
\begin{align}\label{eq:spectral}
\rho_+\pw= \frac{Z}{\pi}\im \frac{-1}{\omega+v|\vp|+i\zeta},
\end{align}
which is depicted in Fig.~\ref{fig:spectle}, where $|\vp|$ is set to zero.
Since the expression of $\zeta$ is not available in the literature,
we simply adopt $\zeta=\cp^4T\ln \cp^{-1}/(2\pi)$ in the figure.

\begin{table}[t]
\begin{center}
\caption{The respective coupling orders of the three fermionic modes.
The momentum of the ultrasoft mode is set to $\ll \cp^2T$, 
while the momenta of the HTL results~\cite{plasmon,scalar} are set to zero.
}
\label{tab:result}
\begin{tabular}{l|c|c}
\hline
 & ultrasoft mode & \parbox{3cm}{normal fermion, \\ antiplasmino} \\ \hline \hline
energy scale & $\ll \cp^2T$  & $\sim \cp T$  \\
damping rate & $\sim \cp^4T\ln \cp^{-1}$ & $\sim \cp^2T$ \\
residue & $\sim \cp^2 $ & $\sim 1$ \\
\hline
\end{tabular} 
\end{center}
\end{table}

\begin{figure*}
\begin{center}
\includegraphics[width=0.6\textwidth]{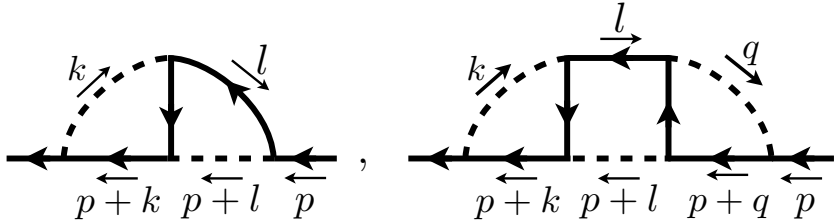}
\caption{Some of the ladder class diagrams.
The solid and dashed line are the dressed propagators of the fermion and the scalar boson, respectively.
}
\label{fig:ladder}
\end{center}
\end{figure*}

\section{Absence of Vertex Correction}
\label{sec:no-vertex}
So far, we have considered the one-loop diagram. 
We need to check that the higher-order loops are suppressed by the coupling constant.
This task would not be straightforward because,  $\delta m^2\sim\cp^2 T^2$
appears in the denominator, as seen in  Eq.~(\ref{eq:one-loop-pc}),
which could make invalid the naive expansion in terms of the coupling constant.
The possible diagrams contributing in the leading order are ladder diagrams 
shown in Fig.~\ref{fig:ladder} because
the fermion-boson pair of the propagators gives a contribution of order $1/\cp^2$, and the vertex gives $\cp^2$.
However, there is a special suppression mechanism in the present case with the scalar coupling.
For example, let us evaluate the first diagram in Fig.~\ref{fig:ladder},
at  small $p$.
The self-energy is evaluated to be
\begin{align}
\begin{split}
\simeq&\frac{\cp^4}{(\delta m^2)^2}\int\frac{d^4k}{(2\pi)^4}K(k)\int\frac{d^4l}{(2\pi)^4}K(l)
\frac{\Slash{k}(\Slash{k}-\Slash{l})\Slash{l}}{(2k\cdot l )}
\\&\times \frac{2p\cdot (k-l)+ 2i(k^0-l^0)\zeta}{\delta m^2} .
\end{split}
\end{align}
Since there are four vertices and two pairs of the propagators whose momenta are almost the same, 
the formula would apparently yield the factor $\cp^4\times(\delta m^2)^{-2}\sim \cp^0$, 
which is the order of unity.
One can easily verify that this order would remain the same in any higher-loop diagram, 
so any ladder diagram seems to contribute in the same leading order as explained.
However, the explicit evaluation of the numerators of the fermion propagators gives
$\Slash{k}(\Slash{k}-\Slash{l})\Slash{l}=\Slash{l}k^2-\Slash{k}l^2$, which
turns out to be of  order $\cp^2$;
this is because the internal particle is almost on-shell, $k^2$, $l^2\sim \cp^2$. 
In the higher-order diagrams such as the second diagram in Fig.~\ref{fig:ladder}, 
the same suppression occurs. Thus, the contribution of 
the ladder diagrams giving a vertex correction 
is absent in the leading order in the scalar coupling, and 
hence the one-loop diagram in Fig.~\ref{fig:oneloop}
 with the dressed propagators solely gives the leading-order contribution
to the self-energy.
We remark that a similar suppression occurs
in the effective three-point-vertex at $p\sim gT$ \cite{scalar}.
It should be emphasized that 
this suppression of the vertex corrections is,  however, not the case in QED/QCD,
 where all the ladder diagrams contribute 
in the leading order and must be summed over~\cite{lebedev2,QED-paper}.

\section{Summary and Concluding Remarks}
\label{sec:summary}

We have investigated the spectral properties of massless fermion 
coupled with a massless scalar boson in the ultrasoft momentum region ($p \lesssim \cp^2T$) 
at high temperature.
We have first indicated that a novel resummed perturbation theory
is needed  beyond the conventional HTL approximation to get a sensible 
spectral properties in the ultrasoft region:
For a consistent calculation in the relevant order of the coupling,
we have shown that the use of the dressed propagators with the asymptotic
masses and the width for the both fermion and boson
is necessary and sufficient; although the vertex corrections
due to ladder diagrams is apparently to contribute in the leading order,
they turn out to be of higher order in the coupling and can be neglected.
We have found that 
the resulting fermion propagator develops a novel pole yielding a peak in the spectral 
function in the ultrasoft region:
Its pole position, width, and the residue are obtained for the first time 
in the present work as a function of the asymptotic masses and damping rates; they
 are summarized and compared with those of the normal-fermion and (anti-)plasmino excitations
obtained in the HTL approximation \cite{HTL,plasmino} 
in Table.~\ref{tab:result}.
Thus, we have established the existence of an ultrasoft fermionic mode 
 at high temperature in the case of scalar coupling for the first time,
although there were suggestions of the existence of such an ultrasoft 
fermionic mode \cite{3peak,mitsutani,shk,lebedev2}. 
We remark that although the present analysis is given for a massless fermion, 
 an ultrasoft fermion excitation should exist
 even for a massive fermion, if the bare mass  is smaller
than $\cp^2T$, which is the smallest scale in our analysis.

We should also emphasize that 
 the existence of a new peak 
at the origin
implies that the fermion spectral function has a three-peak structure,
although the central peak has only a small strength.
The development of such a three-peak structure at intermediate
temperature was suggested 
in the case where the boson is massive irrespective of the type 
of the boson~\cite{3peak,mitsutani,shk}, as mentioned in the Introduction.
In this sense, the three-peak structure of the fermion spectral
function is a robust and persistent phenomenon to be seen 
at intermediate and high $T$, at least in Yukawa theories.

Now let us discuss the physical origin of the ultrasoft mode for
clarifying its possible universality or robustness at high temperature.
As mentioned in the Introduction, an ultrasoft fermionic mode with a vanishing mass can appear
as a phonino associated with the spontaneously broken supersymmetry 
at finite $T$~\cite{phonino,lebedev1,lebedev2}:
In the massive Wess-Zumino model, 
 the supersymmetric cancellation of the fermion mass
is essential to make the phonino excitation~\cite{lebedev1}.
In fact, the massless fermionic mode is realized  as a consequence of an exact cancellation of  
the self-energy with  the finite bare  mass at the vanishing external momentum due to
the spontaneous breaking of supersymmetry.
This is quite in contrast to the case of the Yukawa model dealt in the present work,
where the ultrasoft fermionic modes do {\em not} appear 
when the fermion mass is large.
This means that the ultrasoft fermionic mode like the phonino can appear 
even in a nonsupersymmetric model like the Yukawa model, 
and there the masslessness of the fermion is an essential ingredient 
in realizing it.
Nevertheless it is interesting that such an exotic fermionic mode can  appear
both in supersymmetric and nonsupersymmetric models with quite different mechanisms at high temperature.
For getting further intuition into the possible mechanism for realizing the ultrasoft fermionic
mode in the Yukawa model, it is noted that
Kitazawa {\it{et al.}}  argued that the level repulsion due to the Landau damping
 causes the three-peak structure in Ref.~\cite{3peak},
 though in the case where a massless fermion is coupled with a {\it massive} boson and
thus the fermion-boson mass difference squared $\delta m^2$ is nonzero.
Indeed 
a finite $\delta m^2$ plays an important role in 
the appearance of the ultrasoft mode even when the boson is massless
because it ensures the smoothness of the self-energy at the origin:
As discussed in Introduction, 
the real part of the self-energy vanishes at the origin from the symmetry 
for the massless fermion. 
The residue of the pole is proportional to $(\delta m^2)^2$ and vanishes if $\delta m^2=0$.
In the present work,  the mass difference squared is
a result by taking into account  the effects beyond the HTL approximation
and found to be $\delta m^2\sim \cp^2 T^2$.
The mass difference in turn causes the smallness of the imaginary part:
The imaginary part of the self-energy can originate from 
the boson emission, the Landau damping, and the imaginary part ($\zeta_{f,b}$) 
of the dressed propagators obtained beyond the HTL approximation. 
The former two contributions are found to be zero at the origin \cite{3peak,shk} due to
 the nonzero mass difference or kinematics at the one-loop order.
Therefore, the leading contribution of the imaginary part is solely given by the damping rates
of the hard particles of order $\cp^4T\ln \cp^{-1}$.
As a result, the ultrasoft fermionic mode appears as the sharp peak at the origin 
with a small residue.

Though the present analysis is limited to Yukawa theory with a scalar coupling, 
a similar analysis can be performed 
in other massless fermion-boson system including QED and QCD,
although some complications arise \cite{lebedev2,QED-paper}:
The vertex correction is not negligible in the gauge theories 
in sharp contrast to the Yukawa theory,
so a resummation of the ladder diagram is necessary.
We also note that  the damping rate of the hard fermion is  
 of the order of $\cp^2T\ln \cp^{-1}$ and ``anomalously'' large in QED and QCD \cite{damping-hard-electron}.
Then the imaginary part of the fermion pole in the ultrasoft region should be
larger than that in the Yukawa theory and the nature of the ultrasoft mode 
may be different from that in the present case \cite{QED-paper}.
In QCD, the self-coupling between the gluons should be also taken into account \cite{lebedev2,QED-paper}.
All these matter will be reported in a separate paper \cite {QED-paper}. 

\section*{ACKNOWLEDGMENTS}
Y. H. thanks Robert D. Pisarski for fruitful discussions and comments.
This work was supported by the Grant-in-Aid for the Global COE Program 
``The Next Generation of Physics, Spun from Universality and Emergence'' from the Ministry of Education, 
Culture, Sports, Science and Technology (MEXT) of Japan and by a Grant-in-Aid 
for Scientific Research by the Ministry of Education, Culture, Sports, Science and Technology (MEXT) 
of Japan (No. 20540265).



\begin{thebibliography}{99}
\bibitem{plasmon}
H.~A.~Weldon,
  Phys.\ Rev.\  D {\bf 26} (1982) 1394.

\bibitem{HTL}
J.~Frenkel and J.~C.~Taylor,
  Nucl.\ Phys.\  B {\bf 334} (1990) 199;
    E.~Braaten and R.~D.~Pisarski,
  {\it ibid}. {\bf 339} (1990) 310.


\bibitem{plasmino}
H.~A.~Weldon,
  Phys.\ Rev.\  D {\bf 26} (1982) 2789;
  {\it ibid}. {\bf 40} (1989) 2410.

\bibitem{3peak}
 M.~Kitazawa, T.~Kunihiro and Y.~Nemoto,
  Prog.\ Theor.\ Phys.\  {\bf 117} (2007) 103;
see also 
 M.~Kitazawa, T.~Kunihiro and Y.~Nemoto,
  Phys.\ Lett.\  B {\bf 633} (2006) 269.
%
\bibitem{mitsutani}
 M.~Kitazawa, T.~Kunihiro, K.~Mitsutani and Y.~Nemoto,
  Phys.\ Rev.\  D {\bf 77} (2008) 045034.
%
\bibitem{shk}
 D.~Satow, Y.~Hidaka and T.~Kunihiro,
  Phys.\ Rev.\  D {\bf 83} (2011) 045017.


\bibitem{phonino}
  L.~Girardello, M.~T.~Grisaru and P.~Salomonson,
  Nucl.\ Phys.\  B {\bf 178} (1981) 331;
  D.~Boyanovsky,
  Phys.\ Rev.\  D {\bf 29} (1984) 743;
%
  H.~Aoyama and D.~Boyanovsky,
  Phys.\ Rev.\  D {\bf 30} (1984) 1356;
  R.~Gudmundsdottir and P.~Salomonson,
  Nucl.\ Phys.\  B {\bf 285} (1987) 1.

\bibitem{lebedev1}
 V.~V.~Lebedev and A.~V.~Smilga,
  Nucl.\ Phys.\  B {\bf 318} (1989) 669.
\bibitem{lebedev2}
 V.~V.~Lebedev and A.~V.~Smilga,
  Annals Phys.\  {\bf 202} (1990) 229.
  
  \bibitem{hidaka}
  Y.~Hidaka and T.~Kunihiro,
  Phys.\ Rev.\ D {\bf 83} (2011) 076004.

\bibitem{transport}
  S.~Jeon,
  Phys.\ Rev.\  D {\bf 52} (1995) 3591;
  J.~S.~Gagnon and S.~Jeon,
  Phys.\ Rev.\  D {\bf 75} (2007) 025014
  [Erratum-ibid.\  D {\bf 76} (2007) 089902];
    J.~S.~Gagnon and S.~Jeon,
  Phys.\ Rev.\  D {\bf 76} (2007) 105019.

\bibitem{ultrasoft-am}
 J.~P.~Blaizot and E.~Iancu,
  Nucl.\ Phys.\  B {\bf 570} (2000) 326.


\bibitem{QED-paper}
 Y.~Hidaka and D.~Satow, in preparation.


\bibitem{Keldysh}
  J.~S.~Schwinger,
  J.\ Math.\ Phys.\  {\bf 2} (1961) 407;
  L.~V.~Keldysh,
  Zh.\ Eksp.\ Teor.\ Fiz.\  {\bf 47} (1964) 1515
  [Sov.\ Phys.\ JETP {\bf 20} (1965) 1018].

\bibitem{HTL-resum}
  E.~Braaten and R.~D.~Pisarski,
  Nucl.\ Phys.\  B {\bf 337} (1990) 569.
  
  \bibitem{scalar}
 M.~H.~Thoma,
  Z.\ Phys.\  C {\bf 66} (1995) 491.

\bibitem{dispersion-numerical}
A.~Peshier, K.~Schertler and M.~H.~Thoma,
  Annals Phys.\  {\bf 266} (1998) 162.

\bibitem{Blaizot:2001nr}
  J.~P.~Blaizot and E.~Iancu,
  Phys.\ Rept.\  {\bf 359} (2002) 355.
  
\bibitem{Kapusta}
J.~Kapusta, P.~Lichard and D.~Seibert, Phys. Rev. D {\bf 44}  (1991) 2774 [Erratum-ibid.
{\bf 47} (1991) 4171].
\bibitem{Baier}
R.~Baier, H.~Nakkagawa, A.~Niegawa and K.~Redlich,  Z. Phys. C {\bf 53} (1992) 433.

\bibitem{damping-hard-electron}
 V.~V.~Lebedev and A.~V.~Smilga,
  Phys.\ Lett.\  B {\bf 253} (1991) 231;
  Physica A {\bf 181} (1992) 187.
   J.~P.~Blaizot and E.~Iancu,
  Phys.\ Rev.\ Lett.\  {\bf 76} (1996) 3080;
  Phys.\ Rev.\  D {\bf 55} (1997) 973.
  
\end{thebibliography}
\end{document}